\documentclass[sigconf, nonacm]{acmart}

\newcommand\vldbyear{2025}
\newcommand\vldbworkshop{AIDB}
\newcommand\vldbauthors{\authors}
\newcommand\vldbtitle{\shorttitle}
\newcommand\vldbavailabilityurl{https://github.com/Manuel-Neuer1/Bug_Issue_Link}
\newcommand\vldbpagestyle{plain}

\usepackage{booktabs}
\usepackage{tabularx}
\usepackage{multirow}
\usepackage{tikz}
\usepackage{threeparttable}

\usepackage{xcolor}
\usepackage{tcolorbox}
\usepackage{listings}
\usepackage{makecell}
\definecolor{lightblack}{RGB}{50,50,50}
\definecolor{lightgray}{RGB}{128,128,128}

\usepackage{xspace}

\newcommand{\eg}{{\em e.g., \xspace}}
\newcommand{\esp}{{\em esp., \xspace}}

\newcommand{\squishlist}{
	\begin{list}{$\bullet$}
		{ \setlength{\itemsep}{2pt}      \setlength{\parsep}{3pt}
			\setlength{\topsep}{3pt}       \setlength{\partopsep}{0pt}
			\setlength{\leftmargin}{5.5mm} \setlength{\labelwidth}{1em}
			\setlength{\labelsep}{0.5em} }}
\newcommand{\squishend}{
\end{list}}

\usepackage{fontawesome5}
\newcommand{\bugicon}{{\color{red}\faBug}}

\newcommand*\circled[1]{\tikz[baseline=(char.base)]{
            \node[shape=circle,draw,thick, inner sep=0.8pt] (char) {#1};}}

\begin{document}

\title{LLM-based Dynamic Differential Testing for Database Connectors with Reinforcement Learning-Guided Prompt Selection}
\subtitle{(Extended Abstracts)}

\author{Ce Lyu}
\affiliation{%
  \institution{East China Normal University}
  \country{}
}
\email{51275903097@stu.ecnu.edu.cn}

\author{Minghao Zhao}
\affiliation{%
  \institution{East China Normal University}
  \country{}
}
\email{mhzhao@dase.ecnu.edu.cn}

\author{Yanhao Wang}
\affiliation{%
  \institution{East China Normal University}
  \country{}
}
\email{yhwang@dase.ecnu.edu.cn}

\author{Jie Liang}
\affiliation{%
  \institution{Beihang University}
  \country{}
}
\email{liangjie.mailbox.cn@gmail.com}

\begin{abstract}
Database connectors are critical components enabling applications to interact with underlying database management systems (DBMS), yet their security vulnerabilities often remain overlooked. Unlike traditional software defects, connector vulnerabilities exhibit subtle behavioral patterns and are inherently challenging to detect. 
Besides, nonstandardized implementation of connectors leaves potential risks (a.k.a. unsafe implementations) but is more elusive.
As a result, traditional fuzzing methods are incapable of finding such vulnerabilities. 
Even for LLM-enable test case generation, due to a lack of domain knowledge, they are also incapable of generating test cases that invoke all interface and internal logic of connectors.

In this paper, we propose reinforcement learning (RL)-guided LLM test-case generation for database connector testing.
Specifically, to equip the LLM with sufficient and appropriate domain knowledge, a parameterized prompt template is composed which can be utilized to generate numerous prompts.
Test cases are generated via LLM with a prompt, and are dynamically evaluated through differential testing across multiple connectors.
The testing is iteratively conducted, with each round RL is adopted to select optimal prompt based on prior-round behavioral feedback, so as to maximize control flow coverage.
We implement aforementioned methodology in a practical tool and evaluate it on two widely used JDBC connectors: MySQL Connector/J and OceanBase Connector/J. 
In total, we reported 16 bugs, among them 10 are officially confirmed and the rest are acknowledged as unsafe implementations.

\end{abstract}

\maketitle
\pagestyle{\vldbpagestyle}
\begingroup\small\noindent\raggedright\textbf{VLDB Workshop Reference Format:}\\
\vldbauthors. \vldbtitle. VLDB \vldbyear\ Workshop: \vldbworkshop.\\ 
\endgroup
\begingroup
\renewcommand\thefootnote{}\footnote{\noindent
This work is licensed under the Creative Commons BY-NC-ND 4.0 International License. Visit \url{https://creativecommons.org/licenses/by-nc-nd/4.0/} to view a copy of this license. For any use beyond those covered by this license, obtain permission by emailing \href{mailto:info@vldb.org}{info@vldb.org}. Copyright is held by the owner/author(s). Publication rights licensed to the VLDB Endowment. \\
\raggedright Proceedings of the VLDB Endowment. 
ISSN 2150-8097. \\
}\addtocounter{footnote}{-1}\endgroup

\ifdefempty{\vldbavailabilityurl}{}{
\vspace{.3cm}
\begingroup\small\noindent\raggedright\textbf{VLDB Workshop Artifact Availability:}\\
The source code, data, and/or other artifacts have been made available at \url{https://github.com/Manuel-Neuer1/Bug_Issue_Link}.
\endgroup
}


\section{Introduction}

Database connectors, also known as database drivers,
 serve as crucial middleware that provides standardized interfaces for application-database interactions.
These components translate application API calls into native database commands while transforming query results into application-processable formats. 
Although this abstraction layer significantly enhances development efficiency, any inherent defects in these connectors may propagate system-wide failures.
Thus, it is essential to ensure their reliability and correctness.


Unfortunately, detecting vulnerabilities of database connectors is challenging.
Unlike traditional software defects, connector vulnerabilities exhibit subtle behavioral patterns.
As a result, traditional fuzzing techniques exhibit limited effectiveness in testing database connectors due to their inability to handle protocol-specific syntax and stateful interactions.
Notably, while these fuzzing techniques have proven highly successful for DBMS testing, their effectiveness remains constrained for connectors ~\cite{jiang2023dynsql, rigger2020testing, zhong2020squirrel, fu2022griffin, fu2025understanding,song2025schema, cui2025simple,deng2024coni}.
This is because existing fuzzers primarily generate equivalent SQL queries, whereas the connectors bypass rather than execute them.
Moreover, the {\em static} nature of conventional fuzzing renders it ineffective for comprehensively testing connector logic
-- the generated queries typically exercise only a limited subset of interfaces and achieve insufficient branch coverage. 

What is even worse, certain connector vulnerabilities are scenario-specific or originate from flawed implementation strategies, making such bugs significantly harder to detect.
For example,  
applications often migrate data from one DBMS to another, relying on compatible connectors to handle differences in protocols and SQL dialects. When migrating, differences in connector implementations (even for supposedly compatible ones) can trigger exceptions or exhibit inconsistent behavior on additional connectors, such as data loss, silent rollbacks, or duplicate inserts. These subtle deviations may not generate error messages directly, and thus are difficult to detect. 

connectors are typically expected to comply with standardized interface specifications, such as ODBC (Open Database Connectivity) and JDBC (Java Database Connectivity). However, in practice, some database vendors deviate from -- or even entirely disregard -- these standards, \eg due to compatibility requirements with legacy DBMS.
Such nonstandardized implementation of connectors leaves potential risks (a.k.a. unsafe implementations) but is more elusive.
Besides, the behavior of the connector is highly sensitive to connection property options. These properties can have subtle but critical effects on query execution, transaction processing, and batch processing behavior. In addition, connector behavior is highly dependent on contextual semantics and invocation methods, further increasing the difficulty of writing test cases manually.


\begin{figure*}[t]
    \centering
    \includegraphics[width=\linewidth]{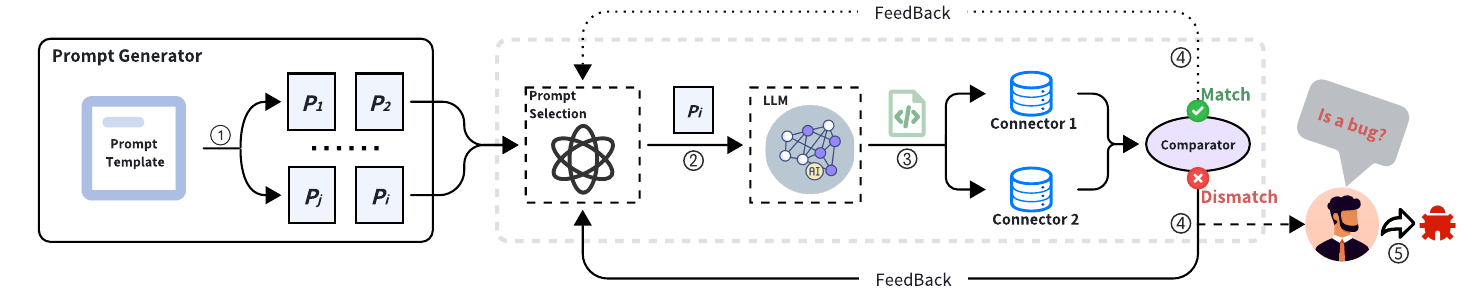}
    \caption{Overall Workflow of LLM-Enhanced DBMS Connector Testing with RL-Guided Prompt Scheduling.\label{fig:arch}}
    \Description{workflow}
\end{figure*}


Recent research has demonstrated the significant potential and initial successes of large language models (LLMs) in software testing applications~\cite{wang2024software, hou2024large, qi2024survey, jiang2024survey}.
However, due to a lack of domain knowledge, it is difficult for the LLMs to generate test cases that invoke all interface and internal logic of connectors.
Besides, static or single prompts often fail to detect deep vulnerabilities, \esp for those defects appears in data transmission among multiple DBMSes. 

In this paper, we propose reinforcement learning (RL)-guided LLM test-case generation for database connector testing.
Specifically, to equip the LLM with sufficient and appropriate domain knowledge, a parameterized prompt template is composed which can be utilized to generate numerous prompts.
Test cases are generated via LLM with a prompt, and are dynamically evaluated through differential testing across multiple connectors.
The testing is iteratively conducted, with each round RL is adopted to select an optimal prompt based on prior-round behavioral feedback, so as to maximize control flow coverage.



By focusing on historically efficient prompts, our approach enables more efficient connector test case generation and was evaluated on MySQL Connector/J and OceanBase Connector/J, where 10 have been confirmed as bugs, and 6 unsafe implementations, as summarized in ~\autoref{tab:bug_summary_detailed}. Some of these bugs have existed for decades without being fixed. These unsafe implementations do not follow the current JDBC specification~\cite{oracle-javadoc-resultset-beforeFirst} due to compatibility with the erroneous behavior of older versions of MySQL Connector/J.
\section{Methodology}

\subsection{Overview}

We propose an architecture as illustrated in \autoref{fig:arch}.
The core idea is to leverage LLMs to automatically generate a diverse suite of test cases.
Beginning with the \emph{Prompt Generator}, it utilizes a structured prompt template to create a set of prompt candidates, which are designed to cover a wide range of connector behaviors (\circled{1}).
Next, an optimal prompt is selected from the candidate set by a RL-Guided strategy, which is then passed to an LLM to generate the test case (\circled{2}).
The generated test case is tested on two compatible connectors for differential testing (\circled{3}).
The \textit{Comparator} then compares the results.
If any inconsistencies are detected, a reward signal is sent back to the \textit{RL Guidance} to reinforce the next iteration of prompt selection (\circled{4}).
Finally, we analyze the cases that have inconsistencies during differential testing, make logic simplifications, and report to the respective development teams (\circled{5}).

\begin{figure}[t]
  \centering
  \includegraphics[width=\linewidth]{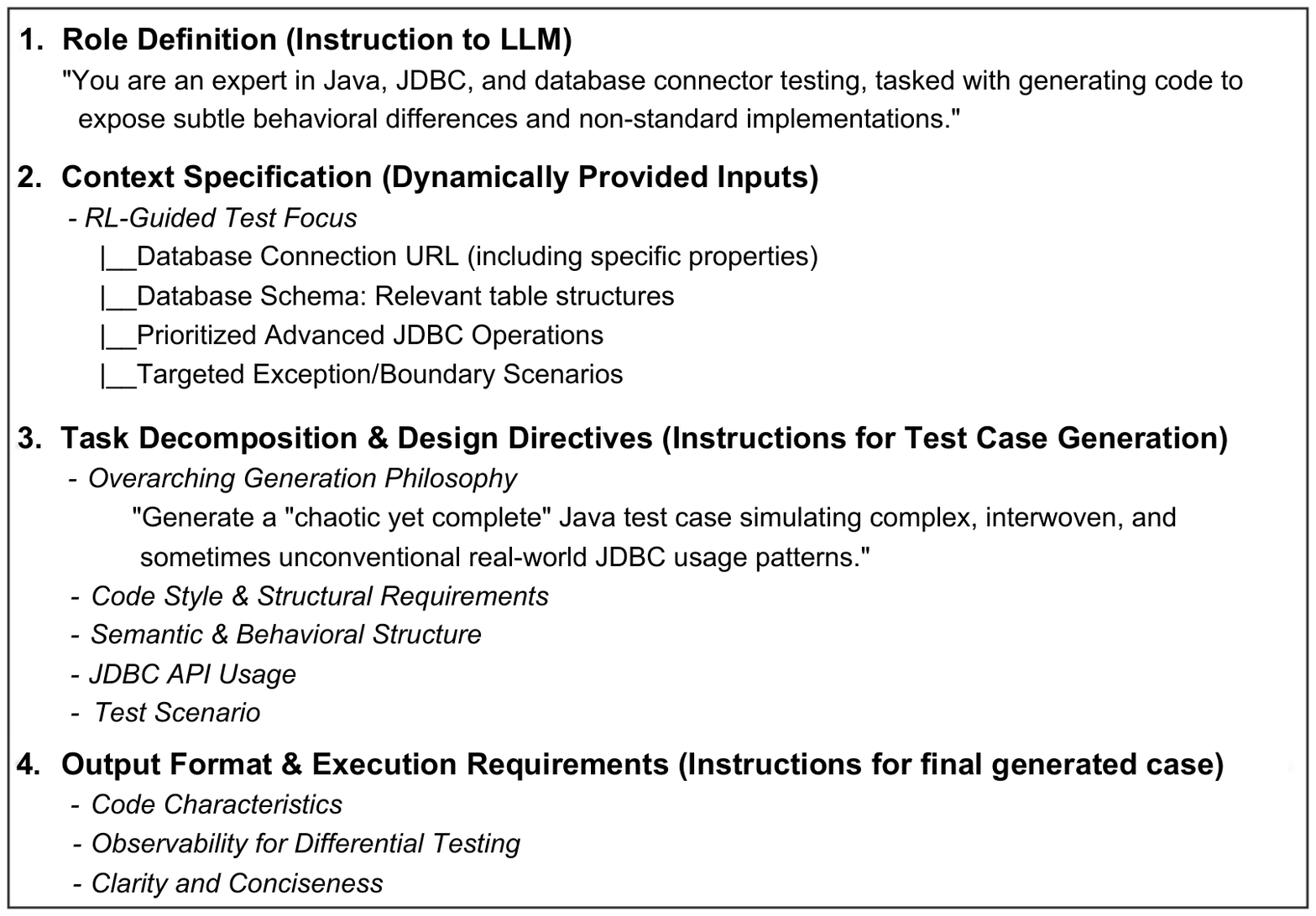}
  \caption{Prompt template for DBMS connector testing.\label{fig:prompt}}
  \Description{prompt}
\end{figure}

Our framework introduces a structured prompt template to instruct the LLM to behave as an expert of DBMS testing.
As detailed in \autoref{fig:prompt}, the template covers four major aspects, namely role definition, dynamic context specification, task decomposition, and output requirements.
This design can more effectively guide the LLM to generate complex and targeted test cases, thereby detecting bugs and unsafe implementations in the connector.

\subsection{Database Connection Property}

The behavior of database connectors can be significantly influenced by their connection-level property parameters, such as \texttt{allowMulti-} \texttt{Queries} and \texttt{rewriteBatchedStatements} in JDBC.
These options affect the internal optimization paths of connectors, query rewriting logic, and exception handling mechanisms.
However, many existing test generators treat the JDBC URL as static, failing to explore the rich behavioral variations induced by different connection property settings.
Thus, they miss a critical axis of behavioral variability.

To address this limitation, guided by domain knowledge, we design a systematic connection property module that explores a diverse set of JDBC parameters during test generation.
Our goal is to maximize the behavioral surface exposed to the downstream connectors under test.

We first identify a set of predefined $m$ JDBC parameters.
Then, we define a property schema $\mathcal{C} = \{c_1, c_2, \dots, c_m\}$, where each $c_j$ corresponds to a boolean or enumerated JDBC parameter.
Each parameter has a well-defined domain.

Rather than exhaustively searching the space $V$ of all possible parameter combinations, i.e.,
\begin{equation}
    \mathcal{V} = \text{Dom}(c_1) \times \text{Dom}(c_2) \times \cdots \times \text{Dom}(c_m),
\end{equation}
we alternatively construct a curated set $\mathbb{S}$ of $k$ representative property subsets, i.e.,
\begin{equation}
    \mathbb{S} = \{ S_1, S_2, \dots, S_k \}, \quad S_i \subseteq \mathcal{V},
\end{equation}
where each $S_i$ is a selected subset of connection properties, which is designed to capture different behaviors.
By running tests on a subset of these connection properties, we can reveal different execution behaviors without exhaustively covering the entire property space. 
This approach allows for extensive and non-exhaustive testing of connector interfaces affected by properties, helping to reveal issues that are missed because of connection properties.

\subsection{RL-Guided Prompt Scheduling}

To maximize the effectiveness of LLM-generated JDBC test cases, we introduce an RL-guided prompt scheduling mechanism, which adaptively selects prompts based on their historical bug-finding performance.
Inspired by the multi-armed bandit (MAB) formulation~\cite{SuttonBarto2018}, we model each prompt template as a distinct arm and apply the Upper Confidence Bound (UCB1) algorithm~\cite{bouneffouf2016finite} to guide prompt selection over multiple testing rounds.

Given a set of prompt templates, we define $\mathcal{P} = \{ P_1, P_2, \ldots, P_N \}$ where \(N \in \mathbb{N}^+\) denotes the total number of available prompt templates.
Our objective is to adaptively identify and prioritize the prompt template ($P_i$) that maximizes the discovery of differential behaviors (indicative of potential bugs) in database connectors.
Each prompt $P_i$ acts as an arm in the MAB setting, with an unknown reward distribution corresponding to the likelihood that test cases generated from $P_i$ expose connector inconsistencies.
We employ the classic UCB1 algorithm to balance exploration and exploitation in prompt selection.
In each round, the algorithm selects the prompt $P_i$ that maximizes
\begin{equation}
    \mu_i + \sqrt{\frac{2\log R}{s_i}}, \quad i \in N.
\end{equation}
For each prompt $P_i$, we maintain the following:
\begin{itemize}
    \item $s_i$: the number of times $P_i$ has been selected;
    \item $\mu_i$: the empirical mean number of output inconsistencies detected by test cases generated from $P_i$;
    \item $R$: the total number of round iterations so far.
\end{itemize}
After generating a test case using the selected prompt, the test is automatically rewritten, compiled, executed, and compared across different JDBC database connectors.
The number of observed behavioral discrepancies serves as a reward signal to update $s_i$ and $\mu_i$.
This closes the feedback loop between prompt selection and the identification of potentially problematic test scenarios, allowing learning-based scheduling to focus on high-impact prompts.

\section{Preliminary Experiments}

To evaluate the effectiveness of our method in discovering bugs and unsafe implementations on database connectors, we answer the following question:
\emph{How does our method perform on real-world database connectors?}

\subsubsection*{\textbf{Tested Database Connectors.}}
We tested two widely used JDBC connectors, namely MySQL Connector/J~\cite{MySQL-connector-j} and OceanBase Connector/J~\cite{oceanbase-connector-j}.
For MySQL Connector/J, we used version 9.2.0 with MySQL 8.0.36.
For OceanBase Connector/J,  we used OceanBase Client 4.2.0 with 5.7.25-OceanBase\_CE-v4.2.1.10.
We use Qwen-plus as the LLM for database connector test case generation.

\begin{table}[t]
    \small
    \centering
    \caption{Number of Bugs and Unsafe Implementations.\label{tab:bug_summary_detailed}}
    \begin{tabular*}{\columnwidth}{@{\extracolsep{\fill}}llc@{}}
        \toprule
        \textbf{Type}          & \textbf{Database Connector} & \textbf{Quantity} \\
        \midrule
        Bugs                   & MySQL / OceanBase       & 7 / 3 \\
        Unsafe Implementations & OceanBase               & 6  \\
        \midrule
        \multicolumn{2}{@{}l}{Total (Bugs + Unsafe Implementations)} & 16 \\
        \bottomrule
    \end{tabular*}
\end{table}

\begin{table*}[t]
\begin{threeparttable}
  \caption{Summary of bugs and unsafe implementations in MySQL and OceanBase Connector/J.}
  \label{tab:commands}
  \small
  \centering
  \setlength{\tabcolsep}{2pt}
  \begin{tabularx}{\linewidth}{ccccX}
    \toprule
    \textbf{ID} & \textbf{Type} & \textbf{Database} & \textbf{Key Aspect / Trigger} & \textbf{Description} \\
    \midrule
    Issue 1 & \faBug & MySQL & Exception Message Error  & Output message error when using \texttt{setMaxRows()} \\
    Issue 2 & \faBug & MySQL & Spec. Violation          & Using \texttt{executeBatch()} to execute a non-DML statement returns an illegal value rather than throws an exception. \\
    Issue 3 & \faBug & MySQL & API Behavior             & Calling \texttt{getHoldability()} is expected to get 1, but actually throws an exception. \\
    Issue 4 & \faBug & MySQL & Config. Interaction      & The \texttt{rewriteBatchedStatements} connection property unexpectedly affects query results following batch inserts. \\
    Issue 5 & \faBug & OceanBase & Resource Management  & \texttt{ResultSet} should be closed, unexpectedly not closed in OceanBase. \\
    Issue 6 & \faBug & MySQL & Config. Interaction      & The \texttt{allowMultiQueries} connection property unexpectedly affects the result of \texttt{getUpdateCounts()} after batch execution. \\
    Issue 7 & \faBug & MySQL & Config. Breaks Atomicity & Atomicity of batch operation is compromised by \texttt{allowMultiQueries}.\\
    Issue 8 & \faBug & OceanBase & Config. Interaction  & The \texttt{rewriteBatchedStatements} connection property unexpectedly affects query results following batch inserts. \\
    Issue 9 & \faBug & OceanBase & Spec. Violation      & Using \texttt{executeBatch()} to execute a non-DML statement returns an illegal value rather than throws an exception. \\
    Issue 10 & \faBug & MySQL & API Behavior            & When sets \texttt{resultSetHoldability} to 2, \texttt{getResultSetHoldability()} unexpectedly returns 1.  \\
    \midrule
    Issue 11 & \faBomb & OceanBase & non-standard       & \multirow{6}{=}{OceanBase compatibility with erroneous behaviors of MySQL JDBC 5.x: \texttt{previous()}, \texttt{first()}, \texttt{afterLast()}, \texttt{absolute()}, \texttt{last()}, and \texttt{beforeFirst()} do not conform to JDBC documentation~\cite{oracle-javadoc-resultset-beforeFirst}, which requires throwing \texttt{SQLException} when called on a \texttt{TYPE\_FORWARD\_ONLY} \texttt{ResultSet}.} \\
    Issue 12 & \faBomb & OceanBase & non standard & \\
    Issue 13 & \faBomb & OceanBase & non-standard & \\
    Issue 14 & \faBomb & OceanBase & non-standard & \\
    Issue 15 & \faBomb & OceanBase & non-standard & \\
    Issue 16 & \faBomb & OceanBase & non-standard & \\
    \bottomrule
  \end{tabularx}
  \begin{tablenotes}[para,flushleft] 
      \footnotesize 
      \item[a] \faBug: Confirmed Bug.
      \item[b] \faBomb: Acknowledged Unsafe Implementation.
      \item[c] Config.: Configuration.
      \item[d] Spec.: specification.
      
    \end{tablenotes}
\end{threeparttable}
\end{table*}

\subsubsection*{\textbf{Confirmed Bugs \& Unsafe Implementations.}}
\autoref{tab:bug_summary_detailed} shows the statistics of our results: seven bugs in MySQL and there bugs and six unsafe implementations in OceanBase.
We briefly describe what triggers each issue and the description in \autoref{tab:commands}.
OceanBase development team provided valuable insight into the emergence of unsafe implementations.
According to an official email response from the OceanBase development team, \emph{``objdbc does not report errors because it is compatible with the erroneous behavior of MySQL-jdbc 5.x. It will be compatible with 8.x in the future''}.
Surprisingly,  Issue 3 remains unfixed in MySQL Connector/J for 17 years!

\subsubsection*{\textbf{Case Study.}}
To concisely demonstrate the detected results, we present two representative case studies: an unsafe implementation in OceanBase Connector/J that deviates from the JDBC specification and a bug in MySQL Connector/J caused by the connection property.
We simplified the logic of test cases to highlight the core issues that trigger errors.

\underline{Case 1:}
Listing~\ref{lst:case1} illustrates an inconsistency between the MySQL and OceanBase connectors.
When invoking \texttt{beforeFirst()} on a \texttt{ResultSet} created with \texttt{TYPE\_FORWARD\_ONLY}, MySQL correctly adheres to the JDBC specification~\cite{oracle-javadoc-resultset-beforeFirst} by triggering a \texttt{SQLException}, while OceanBase executes the same call without exception.
This non-standard implementation may introduce security risks to systems designed to maintain portability between databases.

\begin{lstlisting}[language=Java, caption={MySQL vs. OceanBase: Inconsistent \texttt{beforeFirst()}}, label={lst:case1}, escapeinside={\%*}{*\%}]]
con = DriverManager.%*\textcolor{blue}{getConnection}*\%(url);
stmt = con.%*\textcolor{blue}{createStatement}*\%(ResultSet.TYPE_FORWARD_ONLY);
stmt.%*\textcolor{blue}{executeUpdate}*\%("CREATE TABLE t0 (Id INT);");
rs = stmt.%*\textcolor{blue}{executeQuery}*\%("SELECT Id FROM t0 WHERE Id > 0");
rs.%*\textcolor{blue}{beforeFirst}*\%(); %*\faBomb*\% 
// MySQL triggers SQLException.
%*\textcolor{red}{// OceanBase successfully execute.}%
\end{lstlisting}

\underline{Case 2:}
As shown in Listing~\ref{lst:case2}, which attempts to insert duplicate primary keys (1), (1), (2) into table $t0$, we find an inconsistency bug.
When the connection property \texttt{allowMultiQueries} is enabled, the atomicity of the batch is compromised.
Ideally, the results of the batch operations for primary key conflicts should be consistent, independent of \texttt{allowMultiQueries}.
The MySQL development team has also replied: \emph{``Regardless what the documentation says about the connection property allowMultiQueries, it does affect batched statements''}.

\begin{lstlisting}[language=Java, caption={MySQL: Batch Bug with allowMultiQueries Setting}, label={lst:case2}, escapeinside={\%*}{*\%}]
con = DriverManager.%*\textcolor{blue}{getConnection}*\%(url);
stmt = con.%*\textcolor{blue}{createStatement}*\%();
stmt.%*\textcolor{blue}{execute}*\%("CREATE TABLE t0 (Id INT PRIMARY KEY);");
stmt.%*\textcolor{blue}{addBatch}*\%("INSERT INTO t0 VALUES(1);");
stmt.%*\textcolor{blue}{addBatch}*\%("INSERT INTO t0 VALUES(1);");
stmt.%*\textcolor{blue}{addBatch}*\%("INSERT INTO t0 VALUES(2);");
stmt.%*\textcolor{blue}{executeBatch}*\%(); %*\bugicon*\%
print(); // Assuming prints content of t0
// When allowMultiQueries=true -> print: 1
%*\textcolor{red}{// When allowMultiQueries=false -> print: 1 2 }%
\end{lstlisting}

In summary, both cases confirm that our method can detect bugs and unsafe implementations in database connectors.

\section{Conclusion and Future Work}

In this paper, we studied the problem of testing database connectors.
We proposed a novel framework for this problem based on LLMs with an RL-guided prompt scheduling strategy and identified 10 bugs and 6 unsafe implementations in Oceanbase and MySQL connectors.
As an early-stage study, we aim to answer a central question: \emph{Can our framework find meaningful vulnerabilities in real-world scenarios?}
The results presented a strong affirmative.
In the future, a more comprehensive performance review, baseline comparison, and a deeper analysis of the discovered vulnerabilities will be developed as a more in-depth study.

\bibliographystyle{ACM-Reference-Format}
\bibliography{refs}


\begin{thebibliography}{17}


\ifx \showCODEN    \undefined \def \showCODEN     #1{\unskip}     \fi
\ifx \showDOI      \undefined \def \showDOI       #1{#1}\fi
\ifx \showISBNx    \undefined \def \showISBNx     #1{\unskip}     \fi
\ifx \showISBNxiii \undefined \def \showISBNxiii  #1{\unskip}     \fi
\ifx \showISSN     \undefined \def \showISSN      #1{\unskip}     \fi
\ifx \showLCCN     \undefined \def \showLCCN      #1{\unskip}     \fi
\ifx \shownote     \undefined \def \shownote      #1{#1}          \fi
\ifx \showarticletitle \undefined \def \showarticletitle #1{#1}   \fi
\ifx \showURL      \undefined \def \showURL       {\relax}        \fi
\providecommand\bibfield[2]{#2}
\providecommand\bibinfo[2]{#2}
\providecommand\natexlab[1]{#1}
\providecommand\showeprint[2][]{arXiv:#2}

\bibitem[\protect\citeauthoryear{Bouneffouf}{Bouneffouf}{2016}]%
        {bouneffouf2016finite}
\bibfield{author}{\bibinfo{person}{Djallel Bouneffouf}.} \bibinfo{year}{2016}\natexlab{}.
\newblock \showarticletitle{Finite-time analysis of the multi-armed bandit problem with known trend}. In \bibinfo{booktitle}{\emph{CEC}}. \bibinfo{pages}{2543--2549}.
\newblock


\bibitem[\protect\citeauthoryear{Cui, Dou, Gao, Yang, Zheng, Song, Feng, and Wei}{Cui et~al\mbox{.}}{2025}]%
        {cui2025simple}
\bibfield{author}{\bibinfo{person}{Ziyu Cui}, \bibinfo{person}{Wensheng Dou}, \bibinfo{person}{Yu Gao}, \bibinfo{person}{Rui Yang}, \bibinfo{person}{Yingying Zheng}, \bibinfo{person}{Jiansen Song}, \bibinfo{person}{Yuan Feng}, {and} \bibinfo{person}{Jun Wei}.} \bibinfo{year}{2025}\natexlab{}.
\newblock \showarticletitle{Simple Testing Can Expose Most Critical Transaction Bugs: Understanding and Detecting Write-Specific Serializability Violations in Database Systems}.
\newblock \bibinfo{journal}{\emph{Proc. VLDB Endow.}} \bibinfo{volume}{18}, \bibinfo{number}{8} (\bibinfo{year}{2025}).
\newblock


\bibitem[\protect\citeauthoryear{Deng, Liang, Wu, Fu, Wang, and Jiang}{Deng et~al\mbox{.}}{2024}]%
        {deng2024coni}
\bibfield{author}{\bibinfo{person}{Wenqian Deng}, \bibinfo{person}{Jie Liang}, \bibinfo{person}{Zhiyong Wu}, \bibinfo{person}{Jingzhou Fu}, \bibinfo{person}{Mingzhe Wang}, {and} \bibinfo{person}{Yu Jiang}.} \bibinfo{year}{2024}\natexlab{}.
\newblock \showarticletitle{Coni: Detecting Database Connector Bugs via State-Aware Test Case Generation}. In \bibinfo{booktitle}{\emph{ICSE}}. \bibinfo{pages}{26--37}.
\newblock


\bibitem[\protect\citeauthoryear{Fu, Liang, Wu, Wang, and Jiang}{Fu et~al\mbox{.}}{2022}]%
        {fu2022griffin}
\bibfield{author}{\bibinfo{person}{Jingzhou Fu}, \bibinfo{person}{Jie Liang}, \bibinfo{person}{Zhiyong Wu}, \bibinfo{person}{Mingzhe Wang}, {and} \bibinfo{person}{Yu Jiang}.} \bibinfo{year}{2022}\natexlab{}.
\newblock \showarticletitle{Griffin: Grammar-free DBMS fuzzing}. In \bibinfo{booktitle}{\emph{ASE}}. \bibinfo{pages}{1--12}.
\newblock


\bibitem[\protect\citeauthoryear{Fu, Liang, Wu, Zhao, Li, and Jiang}{Fu et~al\mbox{.}}{2025}]%
        {fu2025understanding}
\bibfield{author}{\bibinfo{person}{Jingzhou Fu}, \bibinfo{person}{Jie Liang}, \bibinfo{person}{Zhiyong Wu}, \bibinfo{person}{Yanyang Zhao}, \bibinfo{person}{Shanshan Li}, {and} \bibinfo{person}{Yu Jiang}.} \bibinfo{year}{2025}\natexlab{}.
\newblock \showarticletitle{Understanding and Detecting SQL Function Bugs: Using Simple Boundary Arguments to Trigger Hundreds of DBMS Bugs}. In \bibinfo{booktitle}{\emph{EuroSys}}. \bibinfo{pages}{1061--1076}.
\newblock


\bibitem[\protect\citeauthoryear{Hou, Zhao, Liu, Yang, Wang, Li, Luo, Lo, Grundy, and Wang}{Hou et~al\mbox{.}}{2024}]%
        {hou2024large}
\bibfield{author}{\bibinfo{person}{Xinyi Hou}, \bibinfo{person}{Yanjie Zhao}, \bibinfo{person}{Yue Liu}, \bibinfo{person}{Zhou Yang}, \bibinfo{person}{Kailong Wang}, \bibinfo{person}{Li Li}, \bibinfo{person}{Xiapu Luo}, \bibinfo{person}{David Lo}, \bibinfo{person}{John Grundy}, {and} \bibinfo{person}{Haoyu Wang}.} \bibinfo{year}{2024}\natexlab{}.
\newblock \showarticletitle{Large language models for software engineering: A systematic literature review}.
\newblock \bibinfo{journal}{\emph{ACM Trans. Softw. Eng. Methodol.}} \bibinfo{volume}{33}, \bibinfo{number}{8} (\bibinfo{year}{2024}), \bibinfo{pages}{1--79}.
\newblock


\bibitem[\protect\citeauthoryear{Jiang, Wang, Shen, Kim, and Kim}{Jiang et~al\mbox{.}}{2024}]%
        {jiang2024survey}
\bibfield{author}{\bibinfo{person}{Juyong Jiang}, \bibinfo{person}{Fan Wang}, \bibinfo{person}{Jiasi Shen}, \bibinfo{person}{Sungju Kim}, {and} \bibinfo{person}{Sunghun Kim}.} \bibinfo{year}{2024}\natexlab{}.
\newblock \showarticletitle{A survey on large language models for code generation}.
\newblock \bibinfo{journal}{\emph{arXiv:2406.00515}} (\bibinfo{year}{2024}).
\newblock


\bibitem[\protect\citeauthoryear{Jiang, Bai, and Su}{Jiang et~al\mbox{.}}{2023}]%
        {jiang2023dynsql}
\bibfield{author}{\bibinfo{person}{Zu-Ming Jiang}, \bibinfo{person}{Jia-Ju Bai}, {and} \bibinfo{person}{Zhendong Su}.} \bibinfo{year}{2023}\natexlab{}.
\newblock \showarticletitle{{DynSQL}: Stateful Fuzzing for Database Management Systems with Complex and Valid {SQL} Query Generation}. In \bibinfo{booktitle}{\emph{USENIX Security}}. \bibinfo{pages}{4949--4965}.
\newblock


\bibitem[\protect\citeauthoryear{{OceanBase Team}}{{OceanBase Team}}{2025}]%
        {oceanbase-connector-j}
\bibfield{author}{\bibinfo{person}{{OceanBase Team}}.} \bibinfo{year}{2025}\natexlab{}.
\newblock \bibinfo{title}{OceanBase Connector/J}.
\newblock \bibinfo{howpublished}{\url{https://github.com/oceanbase/obconnector-j}}.
\newblock


\bibitem[\protect\citeauthoryear{{Oracle}}{{Oracle}}{2014}]%
        {oracle-javadoc-resultset-beforeFirst}
\bibfield{author}{\bibinfo{person}{{Oracle}}.} \bibinfo{year}{2014}\natexlab{}.
\newblock \bibinfo{title}{{ResultSet.beforeFirst() Method}}.
\newblock \bibinfo{howpublished}{\url{https://docs.oracle.com/javase/8/docs/api/java/sql/ResultSet.html\#beforeFirst--}}.
\newblock


\bibitem[\protect\citeauthoryear{{Oracle}}{{Oracle}}{2025}]%
        {MySQL-connector-j}
\bibfield{author}{\bibinfo{person}{{Oracle}}.} \bibinfo{year}{2025}\natexlab{}.
\newblock \bibinfo{title}{MySQL Connector/J}.
\newblock \bibinfo{howpublished}{\url{https://github.com/mysql/mysql-connector-j}}.
\newblock


\bibitem[\protect\citeauthoryear{Qi, Hou, Lin, Bao, and Xu}{Qi et~al\mbox{.}}{2024}]%
        {qi2024survey}
\bibfield{author}{\bibinfo{person}{Fei Qi}, \bibinfo{person}{Yingnan Hou}, \bibinfo{person}{Ning Lin}, \bibinfo{person}{Shanshan Bao}, {and} \bibinfo{person}{Nuo Xu}.} \bibinfo{year}{2024}\natexlab{}.
\newblock \showarticletitle{A Survey of Testing Techniques Based on Large Language Models}. In \bibinfo{booktitle}{\emph{ICCMT}}. \bibinfo{pages}{280--284}.
\newblock


\bibitem[\protect\citeauthoryear{Rigger and Su}{Rigger and Su}{2020}]%
        {rigger2020testing}
\bibfield{author}{\bibinfo{person}{Manuel Rigger} {and} \bibinfo{person}{Zhendong Su}.} \bibinfo{year}{2020}\natexlab{}.
\newblock \showarticletitle{Testing database engines via pivoted query synthesis}. In \bibinfo{booktitle}{\emph{OSDI}}. \bibinfo{pages}{667--682}.
\newblock


\bibitem[\protect\citeauthoryear{Song, Dou, Zheng, Gao, Cui, Wang, and Wei}{Song et~al\mbox{.}}{2025}]%
        {song2025schema}
\bibfield{author}{\bibinfo{person}{Jiansen Song}, \bibinfo{person}{Wensheng Dou}, \bibinfo{person}{Yingying Zheng}, \bibinfo{person}{Yu Gao}, \bibinfo{person}{Ziyu Cui}, \bibinfo{person}{Wei Wang}, {and} \bibinfo{person}{Jun Wei}.} \bibinfo{year}{2025}\natexlab{}.
\newblock \showarticletitle{Detecting Schema-Related Logic Bugs in Relational DBMSs via Equivalent Database Construction}.
\newblock \bibinfo{journal}{\emph{Proc. VLDB Endow.}} \bibinfo{volume}{18}, \bibinfo{number}{7} (\bibinfo{year}{2025}), \bibinfo{pages}{2281--2294}.
\newblock


\bibitem[\protect\citeauthoryear{Sutton and Barto}{Sutton and Barto}{2018}]%
        {SuttonBarto2018}
\bibfield{author}{\bibinfo{person}{Richard~S. Sutton} {and} \bibinfo{person}{Andrew~G. Barto}.} \bibinfo{year}{2018}\natexlab{}.
\newblock \bibinfo{booktitle}{\emph{Reinforcement learning - an introduction, 2nd Edition}}.
\newblock \bibinfo{publisher}{MIT Press}.
\newblock


\bibitem[\protect\citeauthoryear{Wang, Huang, Chen, Liu, Wang, and Wang}{Wang et~al\mbox{.}}{2024}]%
        {wang2024software}
\bibfield{author}{\bibinfo{person}{Junjie Wang}, \bibinfo{person}{Yuchao Huang}, \bibinfo{person}{Chunyang Chen}, \bibinfo{person}{Zhe Liu}, \bibinfo{person}{Song Wang}, {and} \bibinfo{person}{Qing Wang}.} \bibinfo{year}{2024}\natexlab{}.
\newblock \showarticletitle{Software Testing With Large Language Models: Survey, Landscape, and Vision}.
\newblock \bibinfo{journal}{\emph{IEEE Trans. Softw. Eng.}} \bibinfo{volume}{50}, \bibinfo{number}{04} (\bibinfo{year}{2024}), \bibinfo{pages}{911--936}.
\newblock


\bibitem[\protect\citeauthoryear{Zhong, Chen, Hu, Zhang, Lee, and Wu}{Zhong et~al\mbox{.}}{2020}]%
        {zhong2020squirrel}
\bibfield{author}{\bibinfo{person}{Rui Zhong}, \bibinfo{person}{Yongheng Chen}, \bibinfo{person}{Hong Hu}, \bibinfo{person}{Hangfan Zhang}, \bibinfo{person}{Wenke Lee}, {and} \bibinfo{person}{Dinghao Wu}.} \bibinfo{year}{2020}\natexlab{}.
\newblock \showarticletitle{Squirrel: Testing database management systems with language validity and coverage feedback}. In \bibinfo{booktitle}{\emph{CCS}}. \bibinfo{pages}{955--970}.
\newblock


\end{thebibliography}

\end{document}